\documentclass[12pt]{article}%
\usepackage{amsmath}
\usepackage{amsfonts}
\usepackage{amssymb}
\usepackage{graphicx}%
\newtheorem{theorem}{Theorem}

\newtheorem{definition}[theorem]{Definition}

\newtheorem{notation}[theorem]{Notation}

\newtheorem{remark}[theorem]{Remark}

\begin{document}

\title{Solution of the equation $(pu^{\prime})^{\prime}+qu=\omega^{2}u$ by a solution
of the equation $(pu_{0}^{\prime})^{\prime}+qu_{0}=0$}
\author{Vladislav V. Kravchenko\\{\small Department of Mathematics, CINVESTAV del IPN, Unidad Quer\'{e}taro}\\{\small Libramiento Norponiente No. 2000, Fracc. Real de Juriquilla}\\{\small Queretaro, Qro. C.P. 76230 MEXICO}\\{\small e-mail: vkravchenko@qro.cinvestav.mx}}
\maketitle

\begin{abstract}
We give a simple solution of the equation $(pu^{\prime})^{\prime}%
+qu=\omega^{2}u$ whenever a nontrivial solution of $(pu_{0}^{\prime})^{\prime
}+qu_{0}=0$ is known. The method developed for obtaining this result is based
on the theory of pseudoanalytic functions and their relationship with
solutions of the stationary two-dimensional Schr\"{o}dinger equation.
The\ final result, that is the formula for the general solution of the
equation $(pu^{\prime})^{\prime}+qu=\omega^{2}u$ has a simple and easily
verifiable form.

\end{abstract}

\section{Introduction}

The problem of solving the equation
\begin{equation}
\partial_{x}(p(x)\partial_{x}u(x))+q(x)u(x)=\omega^{2}u(x) \label{main}%
\end{equation}
by a known nontrivial solution of the equation
\begin{equation}
\partial_{x}(p(x)\partial_{x}u_{0}(x))+q(x)u_{0}(x)=0 \label{main0}%
\end{equation}
where $p$, $q$, $u$, $u_{0}$ are complex valued functions of the real variable
$x$ and $\omega$ is an arbitrary complex constant is of fundamental importance
due to numerous situations in mathematical physics where it arises. For
example, when the method of separation of variables is applied to the
equation
\[
\operatorname*{div}(P\nabla v)+Qv=0
\]
where $P$ and $Q$ possess some symmetry sufficient for separating variables
very often one can arrive at the equation (\ref{main}), and it is really
desirable to have a possibility to solve only one equation (\ref{main0}) and
to derive from its solution the solution of (\ref{main}). Moreover, in many
important cases the solution of (\ref{main0}) is known. For example, consider
the conductivity equation
\[
\operatorname*{div}(P\nabla v)=0
\]
and suppose, e.g., that $P$ is a function of one Cartesian variable (for a
recent work motivating this example see, e.g., \cite{Demidenko}). Separation
of variables leads to the equation
\[
\partial_{x}(P(x)\partial_{x}u(x))=\omega^{2}u(x)
\]
and the solution of the corresponding equation (\ref{main0}) is given, a
particular nontrivial solution can be chosen as $u_{0}\equiv1$. Thus, to have
a method allowing us to transform $u_{0}$ into $u$ would mean a complete
solution of the original problem.

There are dozens of works dedicated to the construction of zero-energy
solutions of the Schr\"{o}dinger equation (see, e.g.,
\cite{ChadanKobayashi06A}, \cite{ChadanKobayashi06B}). With the aid of the
results of the present paper these solutions can be used for obtaining
solutions for all other values of $\omega$. These are just some immediate
applications of the main result of the present work which can be also applied
in spectral and scattering theories for the Schr\"{o}dinger equation as well
as for studying the Riccati equation.

Here we give a simple and general solution to the problem of solving
(\ref{main}) by a solution of (\ref{main0}). In obtaining it we used some
classical results from pseudoanalytic function theory mainly developed by L.
Bers in fourties and fifties of the last century as well as some recent
results of the author about the relationship between pseudoanalytic functions
and solutions of the stationary two-dimensional Schr\"{o}dinger equation.
Although the main result of this paper was obtained with the aid of the
machinery of pseudoanalytic function theory it can be easily understood,
verified and analized even without those tools. The reader not interested in
the method derived for obtaining it can start reading this paper from
equations (\ref{u1}), (\ref{u2}).

\section{Some definitions and results from pseudoanalytic function theory}

\subsection{The Vekua equation and pseudoanalytic functions}

The main object of study in the pseudoanalytic function theory is the
following equation%
\begin{equation}
W_{\overline{z}}=aW+b\overline{W} \label{Vekua}%
\end{equation}
which is usually called the Vekua equation. Here $a$, $b$ and $W$ are complex
valued functions of a complex variable $z=x+iy$, $W_{\overline{z}}%
=\partial_{\overline{z}}W=\frac{1}{2}(\partial_{x}+i\partial_{y})W$ and
$\overline{W}$ is a complex conjugate of $W$.

Solutions of (\ref{Vekua}) are known as pseudoanalytic (=generalized analytic)
functions. Basic results on them can be found in two well known monographs
\cite{Berskniga} and \cite{Vekua}. In particular, as a part of pseudoanalytic
function theory, L. Bers created a well developed theory of formal powers
which are solutions of (\ref{Vekua}) and locally, near a centre $z_{0}$, they
behave as usual powers $(z-z_{0})^{n}$, $n=0,1,2,\ldots$. Under quite general
conditions he (joint with S. Agmon) proved the expansion theorem which
establishes that any solution of (\ref{Vekua}) in a domain of interest
$\Omega$ can be developed into an infinite series of formal powers and that
this series converges normally (uniformly on any compact subset of $\Omega$).
In posterior works (e.g., \cite{Menke}) stronger results were obtained
guaranteeing the completeness of the infinite system of formal powers in the
space of pseudoanalytic functions in the sense of the $C(\overline{\Omega})$-norm.

In the present work we do not need to consider the general form (\ref{Vekua})
of a Vekua equation. Instead, a very special case is sufficient for our
purposes. In this case the situation with expansion and convergence theorems
is even simpler. We give corresponding details in subsection
\ref{SubsectImpSpecCase}. Here we start with some basic definitions and facts
from \cite{Berskniga}.

\begin{definition}
A pair of complex functions $F$ and $G$ possessing in $\Omega$ partial
derivatives with respect to the real variables $x$ and $y$ is said to be a
generating pair if it satisfies the inequality $\operatorname*{Im}%
(\overline{F}G)>0\qquad$in $\Omega$. The following expressions are known as
characteristic coefficients of the pair $(F,G)$
\[
a_{(F,G)}=-\frac{\overline{F}G_{\overline{z}}-F_{\overline{z}}\overline{G}%
}{F\overline{G}-\overline{F}G},\qquad b_{(F,G)}=\frac{FG_{\overline{z}%
}-F_{\overline{z}}G}{F\overline{G}-\overline{F}G},
\]

\end{definition}%

\[
A_{(F,G)}=-\frac{\overline{F}G_{z}-F_{z}\overline{G}}{F\overline{G}%
-\overline{F}G},\qquad B_{(F,G)}=\frac{FG_{z}-F_{z}G}{F\overline{G}%
-\overline{F}G},
\]
where the subindex $\overline{z}$ or $z$ means the application of
$\partial_{\overline{z}}$ or $\partial_{z}$ respectively.

Every complex function $W$ defined in a subdomain of $\Omega$ admits the
unique representation $W=\phi F+\psi G$ where the functions $\phi$ and $\psi$
are real valued. Sometimes it is convenient to associate with the function $W$
the function $w=\phi+i\psi$. The correspondence between $W$ and $\omega$ is one-to-one.

For $W\in C^{1}(\Omega)$ the $(F,G)$-derivative $\overset{\cdot}{W}%
=\frac{d_{(F,G)}W}{dz}$ exists and has the form
\begin{equation}
\overset{\cdot}{W}=\phi_{z}F+\psi_{z}G=W_{z}-A_{(F,G)}W-B_{(F,G)}\overline{W}
\label{FGder}%
\end{equation}
if and only if
\begin{equation}
\phi_{\overline{z}}F+\psi_{\overline{z}}G=0. \label{phiFpsiG}%
\end{equation}
This last equation can be rewritten in the form (\ref{Vekua}):%
\begin{equation}
W_{\overline{z}}=a_{(F,G)}W+b_{(F,G)}\overline{W}. \label{VekuaFG}%
\end{equation}
Solutions of this equation are called $(F,G)$-pseudoanalytic functions. It is
said that $(F,G)$ is a generating pair corresponding to the Vekua equation
(\ref{VekuaFG}). If $W$ is $(F,G)$-pseudoanalytic, the associated function $w$
is called $(F,G)$-pseudoanalytic of second kind.

\begin{remark}
The functions $F$ and $G$ are $(F,G)$-pseudoanalytic, and $\overset{\cdot}%
{F}\equiv\overset{\cdot}{G}\equiv0$.
\end{remark}

\begin{definition}
\label{DefSuccessor}Let $(F,G)$ and $(F_{1},G_{1})$ - be two generating pairs
in $\Omega$. $(F_{1},G_{1})$ is called \ successor of $(F,G)$ and $(F,G)$ is
called predecessor of $(F_{1},G_{1})$ if%
\[
a_{(F_{1},G_{1})}=a_{(F,G)}\qquad\text{and}\qquad b_{(F_{1},G_{1})}%
=-B_{(F,G)}\text{.}%
\]

\end{definition}

The importance of this definition becomes obvious from the following statement.

\begin{theorem}
\label{ThBersDer}Let $W$ be an $(F,G)$-pseudoanalytic function and let
$(F_{1},G_{1})$ be a successor of $(F,G)$. Then $\overset{\cdot}{W}$ is an
$(F_{1},G_{1})$-pseudoanalytic function.
\end{theorem}

\begin{definition}
\label{DefAdjoint}Let $(F,G)$ be a generating pair. Its adjoint generating
pair $(F,G)^{\ast}=(F^{\ast},G^{\ast})$ is defined by the formulas%
\[
F^{\ast}=-\frac{2\overline{F}}{F\overline{G}-\overline{F}G},\qquad G^{\ast
}=\frac{2\overline{G}}{F\overline{G}-\overline{F}G}.
\]

\end{definition}

The $(F,G)$-integral is defined as follows
\[
\int_{\Gamma}Wd_{(F,G)}z=F(z_{1})\operatorname{Re}\int_{\Gamma}G^{\ast
}Wdz+G(z_{1})\operatorname{Re}\int_{\Gamma}F^{\ast}Wdz
\]
where $\Gamma$ is a rectifiable curve leading from $z_{0}$ to $z_{1}$.

If $W=\phi F+\psi G$ is an $(F,G)$-pseudoanalytic function where $\phi$ and
$\psi$ are real valued functions then
\begin{equation}
\int_{z_{0}}^{z}\overset{\cdot}{W}d_{(F,G)}z=W(z)-\phi(z_{0})F(z)-\psi
(z_{0})G(z), \label{FGAnt}%
\end{equation}
and as $\overset{\cdot}{F}=\overset{}{\overset{\cdot}{G}=}0$, this integral is
path-independent and represents the $(F,G)$-antiderivative of $\overset{\cdot
}{W}$.

\subsection{Generating sequences and Taylor series in formal
powers\label{SubsectGenSeq}}

\begin{definition}
\label{DefSeq}A sequence of generating pairs $\left\{  (F_{m},G_{m})\right\}
$, $m=0,\pm1,\pm2,\ldots$, is called a generating sequence if $(F_{m+1}%
,G_{m+1})$ is a successor of $(F_{m},G_{m})$. If $(F_{0},G_{0})=(F,G)$, we say
that $(F,G)$ is embedded in $\left\{  (F_{m},G_{m})\right\}  $.
\end{definition}

\begin{theorem}
Let \ $(F,G)$ be a generating pair in $\Omega$. Let $\Omega_{1}$ be a bounded
domain, $\overline{\Omega}_{1}\subset\Omega$. Then $(F,G)$ can be embedded in
a generating sequence in $\Omega_{1}$.
\end{theorem}

\begin{definition}
A generating sequence $\left\{  (F_{m},G_{m})\right\}  $ is said to have
period $\mu>0$ if $(F_{m+\mu},G_{m+\mu})$ is equivalent to $(F_{m},G_{m})$
that is their characteristic coefficients coincide.
\end{definition}

Let $W$ be an $(F,G)$-pseudoanalytic function. Using a generating sequence in
which $(F,G)$ is embedded we can define the higher derivatives of $W$ by the
recursion formula%
\[
W^{[0]}=W;\qquad W^{[m+1]}=\frac{d_{(F_{m},G_{m})}W^{[m]}}{dz},\quad
m=1,2,\ldots\text{.}%
\]

\begin{definition}
\label{DefFormalPower}The formal power $Z_{m}^{(0)}(a,z_{0};z)$ with center at
$z_{0}\in\Omega$, coefficient $a$ and exponent $0$ is defined as the linear
combination of the generators $F_{m}$, $G_{m}$ with real constant coefficients
$\lambda$, $\mu$ chosen so that $\lambda F_{m}(z_{0})+\mu G_{m}(z_{0})=a$. The
formal powers with exponents $n=1,2,\ldots$ are defined by the recursion
formula%
\begin{equation}
Z_{m}^{(n)}(a,z_{0};z)=n\int_{z_{0}}^{z}Z_{m+1}^{(n-1)}(a,z_{0};\zeta
)d_{(F_{m},G_{m})}\zeta. \label{recformula}%
\end{equation}

\end{definition}

This definition implies the following properties.

\begin{enumerate}
\item $Z_{m}^{(n)}(a,z_{0};z)$ is an $(F_{m},G_{m})$-pseudoanalytic function
of $z$.

\item If $a^{\prime}$ and $a^{\prime\prime}$ are real constants, then
$Z_{m}^{(n)}(a^{\prime}+ia^{\prime\prime},z_{0};z)=a^{\prime}Z_{m}%
^{(n)}(1,z_{0};z)+a^{\prime\prime}Z_{m}^{(n)}(i,z_{0};z).$

\item The formal powers satisfy the differential relations%
\[
\frac{d_{(F_{m},G_{m})}Z_{m}^{(n)}(a,z_{0};z)}{dz}=nZ_{m+1}^{(n-1)}%
(a,z_{0};z).
\]

\item The asymptotic formulas
\[
Z_{m}^{(n)}(a,z_{0};z)\sim a(z-z_{0})^{n},\quad z\rightarrow z_{0}%
\]
hold.
\end{enumerate}

Assume now that
\begin{equation}
W(z)=\sum_{n=0}^{\infty}Z^{(n)}(a_{n},z_{0};z)\label{series}%
\end{equation}
where the absence of the subindex $m$ means that all the formal powers
correspond to the same generating pair $(F,G),$ and the series converges
uniformly in some neighborhood of $z_{0}$. It can be shown \cite{Berskniga}
that the uniform limit of a series of pseudoanalytic functions is
pseudoanalytic, and that a uniformly convergent series of $(F,G)$%
-pseudoanalytic functions can be $(F,G)$-differentiated term by term. Hence
the function $W$ in (\ref{series}) is $(F,G)$-pseudoanalytic and its $r$th
derivative admits the expansion
\[
W^{[r]}(z)=\sum_{n=r}^{\infty}n(n-1)\cdots(n-r+1)Z_{r}^{(n-r)}(a_{n},z_{0};z).
\]
From this the Taylor formulas for the coefficients are obtained%
\begin{equation}
a_{n}=\frac{W^{[n]}(z_{0})}{n!}.\label{Taylorcoef}%
\end{equation}

\begin{definition}
Let $W(z)$ be a given $(F,G)$-pseudoanalytic function defined for small values
of $\left\vert z-z_{0}\right\vert $. The series%
\begin{equation}
\sum_{n=0}^{\infty}Z^{(n)}(a_{n},z_{0};z) \label{Taylorseries}%
\end{equation}
with the coefficients given by (\ref{Taylorcoef}) is called the Taylor series
of $W$ at $z_{0}$, formed with formal powers.
\end{definition}

\subsection{An important special case\label{SubsectImpSpecCase}}

In the present work in fact we will need the formal powers in the case when
the generating pair has the form
\[
F(x,y)=\frac{\sigma(x)}{\tau(y)}\quad\text{and}\quad G(x,y)=\frac{i\tau
(y)}{\sigma(x)}%
\]
where $\sigma$ and $\tau$ are real-valued functions of their corresponding
variables. For simplicity we assume that $z_{0}=0$ and $F(0)=1$. In this case
(see \cite{Berskniga}) the formal powers are constructed in an elegant manner
as follows. First, denote%

\[
X^{(0)}(x)=\widetilde{X}^{(0)}(x)=Y^{(0)}(y)=\widetilde{Y}^{(0)}(y)=1
\]
and for $n=1,2,...$denote%

\[
X^{(n)}(x)=\left\{
\begin{tabular}
[c]{ll}%
$n%
{\displaystyle\int\limits_{0}^{x}}
X^{(n-1)}(\xi)\frac{d\xi}{\sigma^{2}(\xi)}$ & $\text{for an odd }n$\\
$n%
{\displaystyle\int\limits_{0}^{x}}
X^{(n-1)}(\xi)\sigma^{2}(\xi)d\xi$ & $\text{for an even }n$%
\end{tabular}
\right.
\]

\[
\widetilde{X}^{(n)}(x)=\left\{
\begin{tabular}
[c]{ll}%
$n%
{\displaystyle\int\limits_{0}^{x}}
\widetilde{X}^{(n-1)}(\xi)\sigma^{2}(\xi)d\xi$ & $\text{for an odd }n$\\
$n%
{\displaystyle\int\limits_{0}^{x}}
\widetilde{X}^{(n-1)}(\xi)\frac{d\xi}{\sigma^{2}(\xi)}$ & $\text{for an even
}n$%
\end{tabular}
\ \right.
\]

\[
Y^{(n)}(y)=\left\{
\begin{tabular}
[c]{ll}%
$n%
{\displaystyle\int\limits_{0}^{y}}
Y^{(n-1)}(\eta)\frac{d\eta}{\tau^{2}(\eta)}$ & $\text{for an odd }n$\\
$n%
{\displaystyle\int\limits_{0}^{y}}
Y^{(n-1)}(\eta)\tau^{2}(\eta)d\eta$ & $\text{for an even }n$%
\end{tabular}
\right.
\]

\[
\widetilde{Y}^{(n)}(y)=\left\{
\begin{tabular}
[c]{ll}%
$n%
{\displaystyle\int\limits_{0}^{y}}
\widetilde{Y}^{(n-1)}(\eta)\tau^{2}(\eta)d\eta$ & $\text{for an odd }n$\\
$n%
{\displaystyle\int\limits_{0}^{y}}
\widetilde{Y}^{(n-1)}(\eta)\frac{d\eta}{\tau^{2}(\eta)}$ & $\text{for an even
}n$%
\end{tabular}
\right.
\]
Then for $a=a^{\prime}+ia^{\prime\prime}$ we have%
\[
Z^{(n)}(a,0,z)=\frac{\sigma(x)}{\tau(y)}\operatorname*{Re}\,_{\ast}Z_{{}%
}^{(n)}(a,0,z)+\frac{i\tau(y)}{\sigma(x)}\operatorname*{Im}\,_{\ast}Z_{{}%
}^{(n)}(a,0,z)
\]
where

\begin{align}
_{\ast}Z^{(n)}(a,0,z)  &  =a^{\prime}%
{\displaystyle\sum\limits_{j=0}^{n}}
\binom{n}{j}X^{(n-j)}i^{j}Y^{\left(  j\right)  }\text{\ }\label{Znodd}\\
&  +ia^{\prime\prime}%
{\displaystyle\sum\limits_{j=0}^{n}}
\binom{n}{j}\widetilde{X}^{(n-j)}i^{j}\widetilde{Y}^{\left(  j\right)
}\text{\ \ \ }\quad\text{for an odd }n\nonumber
\end{align}
and%

\begin{align}
_{\ast}Z^{(n)}(a,0,z)  &  =a^{\prime}%
{\displaystyle\sum\limits_{j=0}^{n}}
\binom{n}{j}\widetilde{X}^{(n-j)}i^{j}Y^{\left(  j\right)  }\text{\ }%
\label{Zneven}\\
&  +ia^{\prime\prime}%
{\displaystyle\sum\limits_{j=0}^{n}}
\binom{n}{j}X^{(n-j)}i^{j}\widetilde{Y}^{\left(  j\right)  }\text{\ \ \ \ }%
\quad\text{for an even }n.\nonumber
\end{align}
Consider a rectangular domain containing the origin as an internal point. If
$\sigma$ and $\tau$ are continuously differentiable and bounded together with
$1/\sigma$ and $1/\tau$ on their respective intervals any $(F,G)$%
-pseudoanalytic function can be represented in the form of a Taylor series in
formal powers (\ref{Taylorseries}) with $z_{0}=0$ and the series converges
normally on the domain of interest.

\subsection{The main Vekua equation\label{SectTheMainVekua}}

The equation of the form
\begin{equation}
W_{\overline{z}}=\frac{f_{\overline{z}}}{f}\overline{W} \label{Vekuamain}%
\end{equation}
where $f$ is a nonvanishing continuously differentiable real-valued function
will be called the main Vekua equation. As was shown in \cite{Krpseudoan} this
equation is related to the stationary Schr\"{o}dinger equation much in the
same way as the Cauchy-Riemann system to the Laplace equation. We formulate
two results which will be used throughout the paper.

\begin{theorem}
\cite{Krpseudoan} \label{ThConjugate} Let $W=W_{1}+iW_{2}$ be a solution of
(\ref{Vekuamain}). Then the function $W_{1}$ is a solution of the stationary
Schr\"{o}dinger equation
\begin{equation}
-\Delta W_{1}+q_{1}W_{1}=0\qquad\text{in }\Omega\label{Schr1}%
\end{equation}
with $q_{1}=\Delta f/f,$ and $W_{2}$ is a solution of the associated
stationary Schr\"{o}dinger equation
\begin{equation}
-\Delta W_{2}+q_{2}W_{2}=0\qquad\text{in }\Omega\label{Schr2}%
\end{equation}
where $q_{2}=2(\nabla f)^{2}/f^{2}-q_{1}$ and $(\nabla f)^{2}=f_{x}^{2}%
+f_{y}^{2}$.
\end{theorem}

\begin{notation}
Consider the equation
\begin{equation}
\partial_{\overline{z}}\varphi=\Phi\label{grad}%
\end{equation}
in a whole complex plane or in a convex domain, where $\Phi=\Phi_{1}+i\Phi
_{2}$ is a given complex valued function such that its real part $\Phi_{1}$
and imaginary part $\Phi_{2}$ satisfy the equation
\begin{equation}
\partial_{y}\Phi_{1}-\partial_{x}\Phi_{2}=0, \label{casirot}%
\end{equation}
then as is well known there exist real-valued solutions $\varphi$ to equation
(\ref{grad}) which can be reconstructed up to an arbitrary real constant $c$
in the following way%
\begin{equation}
\varphi(x,y)=2\left(  \int_{x_{0}}^{x}\Phi_{1}(\eta,y)d\eta+\int_{y_{0}}%
^{y}\Phi_{2}(x_{0},\xi)d\xi\right)  +c \label{Antigr}%
\end{equation}
where $(x_{0},y_{0})$ is an arbitrary fixed point in the domain of interest.

By $\overline{A}$ we denote the integral operator in (\ref{Antigr}):%
\[
\overline{A}[\Phi](x,y)=2\left(  \int_{x_{0}}^{x}\Phi_{1}(\eta,y)d\eta
+\int_{y_{0}}^{y}\Phi_{2}(x_{0},\xi)d\xi\right)  +c.
\]
Note that formula (\ref{Antigr}) can be easily extended to any simply
connected domain by considering the integral along an arbitrary rectifiable
curve $\Gamma$ leading from $(x_{0},y_{0})$ to $(x,y)$%
\[
\varphi(x,y)=2\left(  \int_{\Gamma}\Phi_{1}dx+\Phi_{2}dy\right)  +c.
\]

\end{notation}

\begin{theorem}
\label{PrTransform}\cite{Krpseudoan} Let $W_{1}$ be a real valued solution of
(\ref{Schr1}) in a simply connected domain $\Omega$. Then the real valued
function $W_{2},$ solution of (\ref{Schr2}) such that $W=W_{1}+iW_{2}$ is a
solution of (\ref{Vekuamain}), is constructed according to the formula%
\begin{equation}
W_{2}=f^{-1}\overline{A}(if^{2}\partial_{\overline{z}}(f^{-1}W_{1})).
\label{transfDarboux}%
\end{equation}

Given a solution $W_{2}$ of (\ref{Schr2}), the corresponding solution $W_{1}$
of (\ref{Schr1}) such that $W=W_{1}+iW_{2}$ is a solution of (\ref{Vekuamain}%
), is constructed as follows%
\begin{equation}
W_{1}=-f\overline{A}(if^{-2}\partial_{\overline{z}}(fW_{2}%
)).\label{transfDarbouxinv}%
\end{equation}

\end{theorem}

\begin{remark}
Observe that the pair of functions
\begin{equation}
F=f\quad\text{and\quad}G=\frac{i}{f}\label{genpair}%
\end{equation}
is a generating pair for (\ref{Vekuamain}). 
\end{remark}

\section{The main result\label{SectMainresult}}

Consider the equation
\begin{equation}
(-\partial_{x}^{2}+q(x))g(x)=0. \label{Schrwithout}%
\end{equation}
We suppose that $q$ and $g$ are real-valued functions and that on some
interval $I_{x}$ of the independent variable $x$ there exists a bounded
nonvanishing solution $g_{0}\in C^{2}(I_{x})$ such that $1/g_{0}$ is also
bounded. For simplicity we suppose that the interval $I_{x}$ includes the
point $x=0$ and that $g_{0}(0)=1$. Our first goal is to solve the equation%
\[
(-\partial_{x}^{2}+q(x)\pm\omega^{2})u(x)=0
\]
for any real constant $\omega$. We start with the \textquotedblleft%
$+$\textquotedblright-case.

\subsection{The \textquotedblleft$+$\textquotedblright-case}

Consider the equation
\begin{equation}
(-\partial_{x}^{2}+q(x)+\omega^{2})u(x)=0. \label{Schr+}%
\end{equation}
Let us notice that for the equation
\begin{equation}
(-\Delta+q(x)+\omega^{2})U(x,y)=0 \label{biSchr}%
\end{equation}
where $\Delta=\partial_{x}^{2}+\partial_{y}^{2}$ we can immediately propose a
particular solution, e.g.,
\begin{equation}
f(x,y)=g_{0}(x)e^{\omega y}. \label{f}%
\end{equation}
This function does not have zeros on any rectangular domain $\Omega
=I_{x}\times I_{y}$ where $I_{y}$ is an arbitrary finite interval of the
variable $y$. For simplicity we assume that the origin $z=0$ is an internal
point of the domain $\Omega$. According to theorem \ref{PrTransform} any
solution of (\ref{biSchr}) is a real part of a solution of the main Vekua
equation (\ref{Vekuamain}) where $f$ is defined by (\ref{f}). Moreover, a
generating pair for this Vekua equation has the form%
\[
F(x,y)=f(x,y)=g_{0}(x)e^{\omega y}%
\]
and
\[
G(x,y)=i/f(x,y)=ig_{0}^{-1}(x)e^{-\omega y}.
\]
Now using the results from subsection \ref{SubsectImpSpecCase} we can
construct the formal powers corresponding to equation (\ref{Vekuamain}) and to
this generating pair. We have that
\[
Z^{(n)}(a,0,z)=g_{0}(x)e^{\omega y}\operatorname*{Re}\,_{\ast}Z_{{}}%
^{(n)}(a,0,z)+ig_{0}^{-1}(x)e^{-\omega y}\operatorname*{Im}\,_{\ast}Z_{{}%
}^{(n)}(a,0,z)
\]
where $\,_{\ast}Z_{{}}^{(n)}(a,0,z)$ are constructed according to formulas
(\ref{Znodd}) and (\ref{Zneven}) where $\sigma(x)=g_{0}(x)$ and $\tau
(y)=e^{-\omega y}$.

We know that any solution $W$ of (\ref{Vekuamain}) in $\Omega$ can be
represented in the form%
\[
W(z)=%
{\displaystyle\sum\limits_{n=0}^{\infty}}
Z^{(n)}(a_{n},0,z)
\]
and hence any solution $U$ of (\ref{biSchr}) has the form%
\begin{align}
U(x,y)  &  =%
{\displaystyle\sum\limits_{n=0}^{\infty}}
\operatorname*{Re}Z^{(n)}(a_{n},0,z)\label{U}\\
&  =g_{0}(x)e^{\omega y}%
{\displaystyle\sum\limits_{n=0}^{\infty}}
\left(  a_{n}^{\prime}\operatorname*{Re}\,_{\ast}Z^{(n)}(1,0,z)+a_{n}%
^{\prime\prime}\operatorname*{Re}\,_{\ast}Z^{(n)}(i,0,z)\right)  .\nonumber
\end{align}
Observe that solutions of (\ref{Schr+}) are also solutions of (\ref{biSchr}).
Consequently, for any solution $u$ of (\ref{Schr+}) there exists such set of
real numbers $\left\{  a_{n}^{\prime},a_{n}^{\prime\prime}\right\}
_{n=0}^{\infty}$ that
\begin{equation}
u(x)=g_{0}(x)e^{\omega y}%
{\displaystyle\sum\limits_{n=0}^{\infty}}
\left(  a_{n}^{\prime}\operatorname*{Re}\,_{\ast}Z^{(n)}(1,0,z)+a_{n}%
^{\prime\prime}\operatorname*{Re}\,_{\ast}Z^{(n)}(i,0,z)\right)  . \label{u}%
\end{equation}
In other words there exist such sets of coefficients $\left\{  a_{n}^{\prime
},a_{n}^{\prime\prime}\right\}  _{n=0}^{\infty}$ that%
\begin{equation}
\partial_{y}\left(  e^{\omega y}%
{\displaystyle\sum\limits_{n=0}^{\infty}}
\left(  a_{n}^{\prime}\operatorname*{Re}\,_{\ast}Z^{(n)}(1,0,z)+a_{n}%
^{\prime\prime}\operatorname*{Re}\,_{\ast}Z^{(n)}(i,0,z)\right)  \right)
\equiv0. \label{dy=0}%
\end{equation}
Obviously, if this condition is fulfilled, the resulting function (\ref{u}) is
a solution of (\ref{Schr+}).

Let us analyse equation (\ref{dy=0}). First of all we have that for an odd
$n$,%
\[
_{\ast}Z^{(n)}(1,0,z)=%
{\displaystyle\sum\limits_{j=0}^{n}}
\binom{n}{j}X^{(n-j)}i^{j}Y^{\left(  j\right)  },
\]%
\[
_{\ast}Z^{(n)}(i,0,z)=i%
{\displaystyle\sum\limits_{j=0}^{n}}
\binom{n}{j}\widetilde{X}^{(n-j)}i^{j}\widetilde{Y}^{\left(  j\right)  }%
\]
and for an even $n$,
\[
_{\ast}Z^{(n)}(1,0,z)=%
{\displaystyle\sum\limits_{j=0}^{n}}
\binom{n}{j}\widetilde{X}^{(n-j)}i^{j}Y^{\left(  j\right)  },
\]%
\[
_{\ast}Z^{(n)}(i,0,z)=i%
{\displaystyle\sum\limits_{j=0}^{n}}
\binom{n}{j}X^{(n-j)}i^{j}\widetilde{Y}^{\left(  j\right)  }.
\]
Thus, we obtain for an odd $n,$%
\[
\operatorname*{Re}\,_{\ast}Z^{(n)}(1,0,z)=%
{\displaystyle\sum\limits_{\text{even }j=0}^{n}}
\binom{n}{j}X^{(n-j)}i^{j}Y^{\left(  j\right)  },
\]%
\[
\operatorname*{Re}\,_{\ast}Z^{(n)}(i,0,z)=%
{\displaystyle\sum\limits_{\text{odd }j=1}^{n}}
\binom{n}{j}\widetilde{X}^{(n-j)}i^{j+1}\widetilde{Y}^{\left(  j\right)  }%
\]
and for an even $n$,%
\[
\operatorname*{Re}\,_{\ast}Z^{(n)}(1,0,z)=%
{\displaystyle\sum\limits_{\text{even }j=0}^{n}}
\binom{n}{j}\widetilde{X}^{(n-j)}i^{j}Y^{\left(  j\right)  },
\]%
\[
\operatorname*{Re}\,_{\ast}Z^{(n)}(i,0,z)=%
{\displaystyle\sum\limits_{\text{odd }j=1}^{n}}
\binom{n}{j}X^{(n-j)}i^{j+1}\widetilde{Y}^{\left(  j\right)  }.
\]
It is somewhat more convenient for what follows to rewrite these formulas in
the following equivalent form%
\begin{equation}
\left\{
\begin{array}
[c]{c}%
\operatorname*{Re}\,_{\ast}Z^{(n)}(1,0,z)=%
{\displaystyle\sum\limits_{\text{odd }k=1}^{n}}
\binom{n}{k}X^{(k)}i^{n-k}Y^{\left(  n-k\right)  }\\
\operatorname*{Re}\,_{\ast}Z^{(n)}(i,0,z)=%
{\displaystyle\sum\limits_{\text{even }k=0}^{n}}
\binom{n}{k}\widetilde{X}^{(k)}i^{n-k+1}\widetilde{Y}^{\left(  n-k\right)  }%
\end{array}
\right.  \qquad\text{for an odd }n \label{ReZodd}%
\end{equation}
and
\begin{equation}
\left\{
\begin{array}
[c]{c}%
\operatorname*{Re}\,_{\ast}Z^{(n)}(1,0,z)=%
{\displaystyle\sum\limits_{\text{even }k=0}^{n}}
\binom{n}{k}\widetilde{X}^{(k)}i^{n-k}Y^{\left(  n-k\right)  }\\
\operatorname*{Re}\,_{\ast}Z^{(n)}(i,0,z)=%
{\displaystyle\sum\limits_{\text{odd }k=1}^{n}}
\binom{n}{k}X^{(k)}i^{n-k+1}\widetilde{Y}^{\left(  n-k\right)  }%
\end{array}
\right.  \qquad\text{for an even }n \label{ReZeven}%
\end{equation}
We remind that
\[
Y^{(n)}(y)=\left\{
\begin{tabular}
[c]{ll}%
$n%
{\displaystyle\int\limits_{0}^{y}}
Y^{(n-1)}(\eta)e^{2\omega\eta}d\eta$ & $\text{for an odd }n$\\
$n%
{\displaystyle\int\limits_{0}^{y}}
Y^{(n-1)}(\eta)e^{-2\omega\eta}d\eta$ & $\text{for an even }n$%
\end{tabular}
\right.
\]

\[
\widetilde{Y}^{(n)}(y)=\left\{
\begin{tabular}
[c]{ll}%
$n%
{\displaystyle\int\limits_{0}^{y}}
\widetilde{Y}^{(n-1)}(\eta)e^{-2\omega\eta}d\eta$ & $\text{for an odd }n$\\
$n%
{\displaystyle\int\limits_{0}^{y}}
\widetilde{Y}^{(n-1)}(\eta)e^{2\omega\eta}d\eta$ & $\text{for an even }n$%
\end{tabular}
\right.
\]
Thus, from (\ref{ReZodd}) and (\ref{ReZeven}) we have
\begin{equation}
\left\{
\begin{array}
[c]{c}%
\partial_{y}\operatorname*{Re}\,_{\ast}Z^{(n)}(1,0,z)=e^{-2\omega y}%
{\displaystyle\sum\limits_{\text{odd }k=1}^{n}}
(n-k)\binom{n}{k}X^{(k)}i^{n-k}Y^{\left(  n-k-1\right)  }\\
\partial_{y}\operatorname*{Re}\,_{\ast}Z^{(n)}(i,0,z)=e^{-2\omega y}%
{\displaystyle\sum\limits_{\text{even }k=0}^{n}}
(n-k)\binom{n}{k}\widetilde{X}^{(k)}i^{n-k+1}\widetilde{Y}^{\left(
n-k-1\right)  }%
\end{array}
\right.  \qquad\text{for an odd }n \label{dyReZodd}%
\end{equation}
and
\begin{equation}
\left\{
\begin{array}
[c]{c}%
\partial_{y}\operatorname*{Re}\,_{\ast}Z^{(n)}(1,0,z)=e^{-2\omega y}%
{\displaystyle\sum\limits_{\text{even }k=0}^{n}}
(n-k)\binom{n}{k}\widetilde{X}^{(k)}i^{n-k}Y^{\left(  n-k-1\right)  }\\
\partial_{y}\operatorname*{Re}\,_{\ast}Z^{(n)}(i,0,z)=e^{-2\omega y}%
{\displaystyle\sum\limits_{\text{odd }k=1}^{n}}
(n-k)\binom{n}{k}X^{(k)}i^{n-k+1}\widetilde{Y}^{\left(  n-k-1\right)  }%
\end{array}
\right.  \qquad\text{for an even }n>0 \label{dyReZeven}%
\end{equation}
Now returning to equation (\ref{dy=0}) we observe that it is equivalent to the
equation%
\[%
{\displaystyle\sum\limits_{n=0}^{\infty}}
(a_{n}^{\prime}\left(  \omega\operatorname*{Re}\,_{\ast}Z^{(n)}%
(1,0,z)+\partial_{y}\operatorname*{Re}\,_{\ast}Z^{(n)}(1,0,z)\right)
\]%
\[
+a_{n}^{\prime\prime}\left(  \omega\operatorname*{Re}\,_{\ast}Z^{(n)}%
(i,0,z)+\partial_{y}\operatorname*{Re}\,_{\ast}Z^{(n)}(i,0,z)\right)  )=0
\]
from which using (\ref{ReZodd}), (\ref{ReZeven}) and (\ref{dyReZodd}),
(\ref{dyReZeven}) we obtain that (\ref{dy=0}) can be written as follows%
\[
a_{0}^{\prime}\omega+%
{\displaystyle\sum\limits_{\text{even }n=2}^{\infty}}
(a_{n}^{\prime}%
{\displaystyle\sum\limits_{\text{even }k=0}^{n}}
i^{n-k}\binom{n}{k}\widetilde{X}^{(k)}(\omega Y^{\left(  n-k\right)
}+(n-k)Y^{\left(  n-k-1\right)  }e^{-2\omega y})
\]%
\[
+a_{n}^{\prime\prime}%
{\displaystyle\sum\limits_{\text{odd }k=1}^{n}}
i^{n-k+1}\binom{n}{k}X^{(k)}(\omega\widetilde{Y}^{\left(  n-k\right)
}+(n-k)\widetilde{Y}^{\left(  n-k-1\right)  }e^{-2\omega y}))
\]%
\[
+%
{\displaystyle\sum\limits_{\text{odd }n=1}^{\infty}}
(a_{n}^{\prime}%
{\displaystyle\sum\limits_{\text{odd }k=1}^{n}}
i^{n-k}\binom{n}{k}X^{(k)}(\omega Y^{\left(  n-k\right)  }+(n-k)Y^{\left(
n-k-1\right)  }e^{-2\omega y})
\]%
\begin{equation}
+a_{n}^{\prime\prime}%
{\displaystyle\sum\limits_{\text{even }k=0}^{n}}
i^{n-k+1}\binom{n}{k}\widetilde{X}^{(k)}(\omega\widetilde{Y}^{\left(
n-k\right)  }+(n-k)\widetilde{Y}^{\left(  n-k-1\right)  }e^{-2\omega y}))=0.
\label{eqforcoefs}%
\end{equation}
In order that this equality hold identically the expressions corresponding to
different $X^{(n)}$ and $\widetilde{X}^{(n)}$ for all $n$ should vanish
identically. Combining all terms multiplied by $\widetilde{X}^{(0)}$ we obtain
the equation%
\[
a_{0}^{\prime}\omega+%
{\displaystyle\sum\limits_{\text{even }n=2}^{\infty}}
a_{n}^{\prime}i^{n}(\omega Y^{\left(  n\right)  }+nY^{\left(  n-1\right)
}e^{-2\omega y})
\]%
\begin{equation}
+%
{\displaystyle\sum\limits_{\text{odd }n=1}^{\infty}}
a_{n}^{\prime\prime}i^{n+1}(\omega\widetilde{Y}^{\left(  n\right)
}+n\widetilde{Y}^{\left(  n-1\right)  }e^{-2\omega y})=0. \label{firsteq}%
\end{equation}
Gathering all terms multiplied by $X^{(1)}$ we obtain the second equation%
\[%
{\displaystyle\sum\limits_{\text{even }n=2}^{\infty}}
a_{n}^{\prime\prime}i^{n}n(\omega\widetilde{Y}^{\left(  n-1\right)
}+(n-1)\widetilde{Y}^{\left(  n-2\right)  }e^{-2\omega y})
\]%
\[
+%
{\displaystyle\sum\limits_{\text{odd }n=1}^{\infty}}
a_{n}^{\prime}i^{n-1}n(\omega Y^{\left(  n-1\right)  }+(n-1)Y^{\left(
n-2\right)  }e^{-2\omega y})=0
\]
which can be rewritten as follows%
\[
a_{1}^{\prime}\omega+%
{\displaystyle\sum\limits_{\text{even }n=2}^{\infty}}
a_{n+1}^{\prime}i^{n}(n+1)(\omega Y^{\left(  n\right)  }+nY^{\left(
n-1\right)  }e^{-2\omega y})
\]%
\begin{equation}
+%
{\displaystyle\sum\limits_{\text{odd }n=1}^{\infty}}
a_{n+1}^{\prime\prime}i^{n+1}(n+1)(\omega\widetilde{Y}^{\left(  n\right)
}+n\widetilde{Y}^{\left(  n-1\right)  }e^{-2\omega y})=0. \label{secondeq}%
\end{equation}
Gathering all terms multiplied by $\widetilde{X}^{(2)}$, $X^{(3)}$, \ldots\ we
obtain an infinite system of equations which fortunately we do not need to
solve. Here we are reasoning along the following lines. First of all we
observe that if such sets of coefficients $\left\{  a_{n}^{\prime}%
,a_{n}^{\prime\prime}\right\}  _{n=0}^{\infty}$ exist that all the equations
derived from (\ref{eqforcoefs}) are satisfied, they do not depend on functions
$X^{(n)}$ and $\widetilde{X}^{(n)}$ but only on $Y^{\left(  n\right)  }$ and
$\widetilde{Y}^{\left(  n\right)  }$. Thus, they can be constructed
independently of the concrete form of $g_{0}$ and hence of the potential $q$.
Second, we know that such sets of coefficients exist. This is due to our
earlier observation that solutions of (\ref{Schr+}) are also solutions of
(\ref{biSchr}) and hence they can be written in the form (\ref{U}).

These two arguments lead to the following surprising solution. We can take any
$q$, for example, $q\equiv0$ and any pair of independent solutions of the
resulting Schr\"{o}dinger equation (\ref{Schr+}) and to obtain their
corresponding sets of coefficients. These two sets will be universal in the
sense that the general solution of (\ref{Schr+}) for any other $q$ will be
constructed with the aid of this pair of sets of coefficients just changing
the generating function $g_{0}$ and obtaining a corresponding system of
functions $X^{(n)}$ and $\widetilde{X}^{(n)}$. On the first glance this
conclusion can appear against the intuition, nevertheless its more detailed
analysis as well as the final result convince that it is really natural. Thus,
in the next subsection we construct such pair of sets of Taylor coefficients
(in formal powers).

\subsection{Two sets of Taylor coefficients\label{SubsectTwoSets}}

Here we consider the case $q\equiv0$. Then the Schr\"{o}dinger equation
(\ref{Schr+}) becomes%
\begin{equation}
(-\partial_{x}^{2}+\omega^{2})u(x)=0. \label{Helmh}%
\end{equation}
Note that the corresponding equation (\ref{Schrwithout}) has the form
$\partial_{x}^{2}g(x)=0$ and possesses a suitable particular solution
satisfying all the requirements (see the beginning of section
\ref{SectMainresult}) $g_{0}\equiv1$. Then $f=e^{\omega y}$ and the main Vekua
equation in this case has the form%
\begin{equation}
W_{\overline{z}}=\frac{i\omega}{2}\overline{W}. \label{Vekuamain1}%
\end{equation}
Let us take two independent solutions of (\ref{Helmh}) $u^{+}(x)=e^{\omega x}$
and $u^{-}(x)=e^{-\omega x}$. First, we obtain the set of coefficients
$\left\{  a_{n}^{\prime},a_{n}^{\prime\prime}\right\}  _{n=0}^{\infty}$ for
the function $u^{+}$. The first step consists in constructing the
corresponding conjugate metaharmonic function $v^{+}$ (see theorem
\ref{PrTransform})%
\[
v^{+}=e^{-\omega y}\overline{A}(ie^{2\omega y}\partial_{\overline{z}}%
e^{\omega(x-y)})=e^{-\omega y}\overline{A}(\frac{\omega}{2}(1+i)e^{\omega
(x+y)}).
\]
We have
\[
\overline{A}(\frac{\omega}{2}(1+i)e^{\omega(x+y)})=e^{\omega(x+y)}+c.
\]
We choose $c=0$, then $v^{+}=e^{\omega x}$. Thus, one of the solutions of the
main Vekua equation (\ref{Vekuamain1}) such that $u^{+}=e^{\omega x}$ is its
real part has the form%
\begin{equation}
W^{+}=(1+i)e^{\omega x}. \label{W}%
\end{equation}
Now, in order to construct its corresponding Taylor coefficients (in formal
powers) we notice that $A_{(F,G)}=0$ and $B_{(F,G)}=-i\omega/2$ (here
$F=e^{\omega y}$ and $G=ie^{-\omega y}$). Thus, the operation of the
$(F,G)$-derivative has the form%
\[
\overset{\cdot}{W}=W_{z}+\frac{i\omega}{2}\overline{W}.
\]
For the function (\ref{W}) we have%
\[
\overset{\cdot}{W^{+}}=\omega(1+i)e^{\omega x},\qquad\overset{\cdot\cdot
}{W^{+}}=\omega^{2}(1+i)e^{\omega x},\ldots
\]
and it is easy to see that the $n$-th $(F,G)$-derivative of $W^{+}$ has the
form
\[
W^{+[n]}=\omega^{n}W^{+}.
\]
We obtain that the Taylor coefficients (in formal powers) of the function
(\ref{W}) at the origin have the following simple form%
\begin{equation}
a_{n}^{+}=\frac{\omega^{n}}{n!}(1+i). \label{an+}%
\end{equation}
In a similar way we study the case of the function $u^{-}$. The corresponding
pseudoanalytic function $W^{-}$ has the form $W^{-}=(1-i)e^{-\omega x},$ and
the corresponding Taylor coefficients at the origin are as follows%
\begin{equation}
a_{n}^{-}=\frac{\left(  -\omega\right)  ^{n}}{n!}(1-i). \label{an-}%
\end{equation}
Let us notice that from the fulfillment of (\ref{firsteq}) with the
coefficients of the form (\ref{an+}) or (\ref{an-}) there follows the
fulfillment of (\ref{secondeq}) and of all subsequent equations corresponding
to $\widetilde{X}^{(2)}$, $X^{(3)}$, etc. This is because of the fact that
$a_{n+1}^{\pm}=\frac{\pm\omega}{n+1}a_{n}^{\pm}$.

\subsection{General solution of (\ref{Schr+})}

Now with the aid of the sets of coefficients (\ref{an+}) and (\ref{an-}) we
proceed in obtaining the general solution of (\ref{Schr+}) with any potential
$q$ for which a solution $g_{0}$ of (\ref{Schrwithout}) satisfying the nonzero
and boundedness requirements exists. From (\ref{u}) we have that the general
solution of (\ref{Schr+}) has the form%
\[
u=c_{1}u_{1}+c_{2}u_{2}%
\]
where $c_{1}$ and $c_{2}$ are arbitrary real constants and $u_{1}$, $u_{2}$
are defined as follows
\[
u_{1}(x)=g_{0}(x)e^{\omega y}%
{\displaystyle\sum\limits_{n=0}^{\infty}}
\frac{\omega^{n}}{n!}\left(  \operatorname*{Re}\,_{\ast}Z^{(n)}%
(1,0,z)+\operatorname*{Re}\,_{\ast}Z^{(n)}(i,0,z)\right)
\]
and
\[
u_{2}(x)=g_{0}(x)e^{\omega y}%
{\displaystyle\sum\limits_{n=0}^{\infty}}
\frac{\left(  -\omega\right)  ^{n}}{n!}\left(  \operatorname*{Re}\,_{\ast
}Z^{(n)}(1,0,z)-\operatorname*{Re}\,_{\ast}Z^{(n)}(i,0,z)\right)
\]
which according to (\ref{ReZodd}) and (\ref{ReZeven}) can be written in the
following form%
\begin{align*}
u_{1}(x)  &  =g_{0}(x)e^{\omega y}(%
{\displaystyle\sum\limits_{\text{even }n=0}^{\infty}}
\frac{\omega^{n}}{n!}(%
{\displaystyle\sum\limits_{\text{even }k=0}^{n}}
i^{n-k}\binom{n}{k}\widetilde{X}^{(k)}Y^{\left(  n-k\right)  }\\
&  +%
{\displaystyle\sum\limits_{\text{odd }k=1}^{n}}
i^{n-k+1}\binom{n}{k}X^{(k)}\widetilde{Y}^{\left(  n-k\right)  })\\
&  +%
{\displaystyle\sum\limits_{\text{odd }n=1}^{\infty}}
\frac{\omega^{n}}{n!}(%
{\displaystyle\sum\limits_{\text{odd }k=1}^{n}}
i^{n-k}\binom{n}{k}X^{(k)}Y^{\left(  n-k\right)  }\\
&  +%
{\displaystyle\sum\limits_{\text{even }k=0}^{n}}
i^{n-k+1}\binom{n}{k}\widetilde{X}^{(k)}\widetilde{Y}^{\left(  n-k\right)  }))
\end{align*}
and%
\begin{align*}
u_{2}(x)  &  =g_{0}(x)e^{\omega y}(%
{\displaystyle\sum\limits_{\text{even }n=0}^{\infty}}
\frac{\omega^{n}}{n!}(%
{\displaystyle\sum\limits_{\text{even }k=0}^{n}}
i^{n-k}\binom{n}{k}\widetilde{X}^{(k)}Y^{\left(  n-k\right)  }\\
&  -%
{\displaystyle\sum\limits_{\text{odd }k=1}^{n}}
i^{n-k+1}\binom{n}{k}X^{(k)}\widetilde{Y}^{\left(  n-k\right)  })\\
&  -%
{\displaystyle\sum\limits_{\text{odd }n=1}^{\infty}}
\frac{\omega^{n}}{n!}(%
{\displaystyle\sum\limits_{\text{odd }k=1}^{n}}
i^{n-k}\binom{n}{k}X^{(k)}Y^{\left(  n-k\right)  }\\
&  -%
{\displaystyle\sum\limits_{\text{even }k=0}^{n}}
i^{n-k+1}\binom{n}{k}\widetilde{X}^{(k)}\widetilde{Y}^{\left(  n-k\right)
})).
\end{align*}
As we know that both expressions are independent of $y$ in order to simplify
them we can substitute any value of $y$. Of course, the easiest way is to
substitute $y=0$ because by definition all $Y^{(n)}(0)$ and $\widetilde
{Y}^{(n)}(0)$ for $n\geq1$ are equal to zero, and $Y^{(0)}(0)=\widetilde
{Y}^{(0)}(0)=1$. Thus, finally we obtain%
\begin{equation}
u_{1}(x)=g_{0}(x)(%
{\displaystyle\sum\limits_{\text{even }n=0}^{\infty}}
\frac{\omega^{n}}{n!}\widetilde{X}^{(n)}+%
{\displaystyle\sum\limits_{\text{odd }n=1}^{\infty}}
\frac{\omega^{n}}{n!}X^{(n)}) \label{u1}%
\end{equation}
and
\begin{equation}
u_{2}(x)=g_{0}(x)(%
{\displaystyle\sum\limits_{\text{even }n=0}^{\infty}}
\frac{\omega^{n}}{n!}\widetilde{X}^{(n)}-%
{\displaystyle\sum\limits_{\text{odd }n=1}^{\infty}}
\frac{\omega^{n}}{n!}X^{(n)}). \label{u2}%
\end{equation}
where
\begin{equation}
\widetilde{X}^{(0)}\equiv1,\quad X^{(0)}\equiv1, \label{X1}%
\end{equation}

\begin{equation}
\widetilde{X}^{(n)}(x)=\left\{
\begin{tabular}
[c]{ll}%
$n%
{\displaystyle\int\limits_{0}^{x}}
\widetilde{X}^{(n-1)}(\xi)g_{0}^{2}(\xi)d\xi$ & $\text{for an odd }n$\\
$n%
{\displaystyle\int\limits_{0}^{x}}
\widetilde{X}^{(n-1)}(\xi)g_{0}^{-2}(\xi)d\xi$ & $\text{for an even }n$%
\end{tabular}
\ \right.  \label{X2}%
\end{equation}

\begin{equation}
X^{(n)}(x)=\left\{
\begin{tabular}
[c]{ll}%
$n%
{\displaystyle\int\limits_{0}^{x}}
X^{(n-1)}(\xi)g_{0}^{-2}(\xi)d\xi$ & $\text{for an odd }n$\\
$n%
{\displaystyle\int\limits_{0}^{x}}
X^{(n-1)}(\xi)g_{0}^{2}(\xi)d\xi$ & $\text{for an even }n$%
\end{tabular}
\ \right.  \label{X3}%
\end{equation}
In the next subsection we validate this result by a direct substitution into
equation (\ref{Schr+}).

\subsection{Validating the result}

In order to substitute (\ref{u1}) and (\ref{u2}) or equivalently
\[
v_{1}(x)=g_{0}(x)%
{\displaystyle\sum\limits_{\text{even }n=0}^{\infty}}
\frac{\omega^{n}}{n!}\widetilde{X}^{(n)}%
\]
and
\[
v_{2}(x)=g_{0}(x)%
{\displaystyle\sum\limits_{\text{odd }n=1}^{\infty}}
\frac{\omega^{n}}{n!}X^{(n)}%
\]
into equation (\ref{Schr+}) we first make some helpful observations.

It is well known that a nonvanishing solution $g_{0}$ of (\ref{Schrwithout})
allows us to factorize the Schr\"{o}dinger operator as follows%
\begin{equation}
\partial_{x}^{2}-q(x)=\left(  \partial_{x}+\frac{g_{0}^{\prime}}{g_{0}%
}\right)  \left(  \partial_{x}-\frac{g_{0}^{\prime}}{g_{0}}\right)
.\label{fact}%
\end{equation}
The first order operators in their turn can be factorized as well, so we
obtain%
\[
\partial_{x}^{2}-q=g_{0}^{-1}\partial_{x}g_{0}^{2}\partial_{x}g_{0}^{-1}.
\]
Now let us consider $v_{1}$. By definition, for an even $n$ we have%
\[
\widetilde{X}^{(n)}(x)=n%
{\displaystyle\int\limits_{0}^{x}}
\widetilde{X}^{(n-1)}(\xi)\frac{d\xi}{g_{0}^{2}(\xi)}.
\]
Thus, application of the operator $\partial_{x}^{2}-q$ to $g_{0}\widetilde
{X}^{(n)}$ for an even $n$ and $n\geq2$ (for $n=0$ the result is zero) gives
us%
\begin{align*}
\left(  \partial_{x}^{2}-q\right)  \left(  g_{0}\widetilde{X}^{(n)}\right)
&  =g_{0}^{-1}\partial_{x}g_{0}^{2}\partial_{x}\widetilde{X}^{(n)}=ng_{0}%
^{-1}\partial_{x}\widetilde{X}^{(n-1)}\\
&  =(n-1)ng_{0}\widetilde{X}^{(n-2)}.
\end{align*}
Then
\begin{align*}
\left(  \partial_{x}^{2}-q\right)  v_{1} &  =g_{0}%
{\displaystyle\sum\limits_{\text{even }n=2}^{\infty}}
\frac{\omega^{n}}{\left(  n-2\right)  !}\widetilde{X}^{(n-2)}\\
&  =\omega^{2}g_{0}%
{\displaystyle\sum\limits_{\text{even }n=0}^{\infty}}
\frac{\omega^{n}}{n!}\widetilde{X}^{(n)}=\omega^{2}v_{1}.
\end{align*}
In a similar way one can verify that $v_{2}$ is a solution of (\ref{Schr+}) as
well. Note that according to the general result formulated in subsection
\ref{SubsectImpSpecCase} both series in $v_{1}$ and $v_{2}$ are uniformly
convergent on the interval $I_{x}$. This fact can be quite easily verified as
well by estimating the integrals in $\widetilde{X}^{(n)}$ and $X^{(n)}$ by the
supremum of the functions $g_{0}^{2}$ and $g_{0}^{-2}$ multiplied by
successive antiderivatives of $x$.

\subsection{The \textquotedblleft-\textquotedblright\ case}

Consider the equation
\begin{equation}
(-\partial_{x}^{2}+q(x)-\omega^{2})u(x)=0 \label{Schr-}%
\end{equation}
and the corresponding two-dimensional equation
\begin{equation}
(-\Delta+q(x)-\omega^{2})U(x,y)=0. \label{biSchr-}%
\end{equation}
Its particular solution can be chosen as
\[
f(x,y)=g_{0}(x)\cos\omega y
\]
which is different from zero on the rectangular domain $\Omega=I_{x}%
\times(-\frac{\pi}{2\omega},\frac{\pi}{2\omega})$. In order to obtain the
general solution of (\ref{Schr-}) in fact we should only obtain two sets of
Taylor coefficients as in subsection \ref{SubsectTwoSets}. For this, once more
we take $q\equiv0$ and consider two linearly independent solutions of the
equation $(\partial_{x}^{2}+\omega^{2})u(x)=0$, $u^{+}(x)=\cos\omega x$ and
$u^{-}(x)=\sin\omega x$. The next step is to construct $v^{+}$ and $v^{-}$. We
have%
\[
v^{+}=\frac{1}{\cos\omega y}\overline{A}\left(  i\cos^{2}\omega y\partial
_{\overline{z}}\left(  \frac{\cos\omega x}{\cos\omega y}\right)  \right)
=-\sin\omega x\tan\omega y
\]
(we have fixed the arbitrary constant as zero). Thus,
\[
W^{+}=\cos\omega x-i\sin\omega x\tan\omega y.
\]
In a similar way we obtain%
\[
W^{-}=\sin\omega x+i\cos\omega x\tan\omega y.
\]
Noting that the definition of the $(F,G)$-derivative in this case has the form%
\[
\overset{\cdot}{W}=W_{z}-\frac{i\omega}{2}\tan\omega y\overline{W}%
\]
we obtain the following relations $\overset{\cdot}{W^{+}}=-\omega W^{-}$ and
$\overset{\cdot}{W}\,^{-}=\omega W^{+}$ and hence the following formulas for
the corresponding Taylor coefficients in formal powers in the origin%
\[
a_{n}^{+}=\frac{(i\omega)^{n}}{n!}\text{ for an even }n\quad\text{and}\quad
a_{n}^{+}=0\text{ for an odd }n\text{,}%
\]%
\[
a_{n}^{-}=0\text{ for an even }n\quad\text{and}\quad a_{n}^{-}=\frac
{-i(i\omega)^{n}}{n!}\text{ for an odd }n\text{.}%
\]
Thus we arrive at the following general solution of equation (\ref{Schr-}) for
any potential $q$ admitting a particular solution $g_{0}$ with the described
above properties%
\[
u=c_{1}u_{1}+c_{2}u_{2}%
\]
with
\[
u_{1}(x)=g_{0}(x)%
{\displaystyle\sum\limits_{\text{even }n=0}^{\infty}}
\frac{(i\omega)^{n}}{n!}\widetilde{X}^{(n)}%
\]
and
\[
u_{2}(x)=g_{0}(x)%
{\displaystyle\sum\limits_{\text{odd }n=1}^{\infty}}
\frac{i(i\omega)^{n}}{n!}X^{(n)}%
\]
where $X^{(n)}$ and $\widetilde{X}^{(n)}$ are defined by (\ref{X1})-(\ref{X3}).

\subsection{Complex potential}

It is clear that the results obtained in the preceding subsections remain
valid in the case of a complex valued potential $q$ and an arbitrary complex
number $\omega$. Consider the equation
\begin{equation}
(-\partial_{x}^{2}+q(x)+\omega^{2})u(x)=0 \label{Schrmain}%
\end{equation}
where $q$ and $u$ are complex valued and $\omega$ is any complex number. We
assume that $g_{0}$ is a nonvanishing solution of the equation $(-\partial
_{x}^{2}+q(x))g_{0}=0$ satisfying the boundedness requirements. Then the
general solution of (\ref{Schrmain}) has the form
\begin{equation}
u=c_{1}u_{1}+c_{2}u_{2} \label{gensol}%
\end{equation}
where $c_{1}$ and $c_{2}$ are arbitrary complex constants and $u_{1}$, $u_{2}$
are defined as follows
\begin{equation}
u_{1}=g_{0}%
{\displaystyle\sum\limits_{\text{even }n=0}^{\infty}}
\frac{\omega^{n}}{n!}\widetilde{X}^{(n)} \label{gensol1}%
\end{equation}
and
\begin{equation}
u_{2}=g_{0}%
{\displaystyle\sum\limits_{\text{odd }n=1}^{\infty}}
\frac{\omega^{n}}{n!}X^{(n)} \label{gensol2}%
\end{equation}
where as before $X^{(n)}$ and $\widetilde{X}^{(n)}$ are defined by
(\ref{X1})-(\ref{X3}).

\subsection{Equation $(pu^{\prime})^{\prime}+qu=\omega^{2}u$}

Having obtained the general solution of (\ref{Schrmain}) it is easy to obtain
the\ general solution of the more general equation%
\begin{equation}
\partial_{x}(p\partial_{x}u)+qu=\omega^{2}u\label{Schrgen}%
\end{equation}
where $p\in C^{2}(I_{x})$ is a nonvanishing complex valued function, $q$ and
$u$ satisfy conditions from the preceding subsection. We assume that
\begin{equation}
\partial_{x}(p\partial_{x}g_{0})+qg_{0}=0\label{eqg0}%
\end{equation}
and observe that the following factorization holds
\[
(\partial_{x}p\partial_{x}+q)u=p^{1/2}\left(  \partial_{x}+\frac{g^{\prime}%
}{g}\right)  \left(  \partial_{x}-\frac{g^{\prime}}{g}\right)  (p^{1/2}%
u)=g_{0}^{-1}\partial_{x}\left(  g^{2}\partial_{x}\left(  g_{0}^{-1}u\right)
\right)
\]
where $g=p^{1/2}g_{0}$ and by analogy with (\ref{gensol})-(\ref{gensol2}) we
obtain the following solution of (\ref{Schrgen})%
\begin{equation}
u=c_{1}u_{1}+c_{2}u_{2}\label{genmain}%
\end{equation}
where%
\begin{equation}
u_{1}=g_{0}%
{\displaystyle\sum\limits_{\text{even }n=0}^{\infty}}
\frac{\omega^{n}}{n!}\widetilde{X}^{(n)}\label{genmain1}%
\end{equation}
and
\begin{equation}
u_{2}=g_{0}%
{\displaystyle\sum\limits_{\text{odd }n=1}^{\infty}}
\frac{\omega^{n}}{n!}X^{(n)}\label{genmain2}%
\end{equation}
where the definition of $X^{(n)}$ and $\widetilde{X}^{(n)}$ is slightly
modified:%
\begin{equation}
\widetilde{X}^{(0)}\equiv1,\quad X^{(0)}\equiv1,\label{Xgen1}%
\end{equation}

\begin{equation}
\widetilde{X}^{(n)}(x)=\left\{
\begin{tabular}
[c]{ll}%
$n%
{\displaystyle\int\limits_{0}^{x}}
\widetilde{X}^{(n-1)}(\xi)g_{0}^{2}(\xi)d\xi$ & $\text{for an odd }n$\\
$n%
{\displaystyle\int\limits_{0}^{x}}
\widetilde{X}^{(n-1)}(\xi)g^{-2}(\xi)d\xi$ & $\text{for an even }n$%
\end{tabular}
\ \right.  \label{Xgen2}%
\end{equation}

\begin{equation}
X^{(n)}(x)=\left\{
\begin{tabular}
[c]{ll}%
$n%
{\displaystyle\int\limits_{0}^{x}}
X^{(n-1)}(\xi)g^{-2}(\xi)d\xi$ & $\text{for an odd }n$\\
$n%
{\displaystyle\int\limits_{0}^{x}}
X^{(n-1)}(\xi)g_{0}^{2}(\xi)d\xi$ & $\text{for an even }n$%
\end{tabular}
\ \ \right.  \label{Xgen3}%
\end{equation}
Let us verify that $u_{1}$ is indeed a solution of (\ref{Schrgen}). We have
\begin{align*}
(\partial_{x}p\partial_{x}+q)u_{1} &  =g_{0}^{-1}\partial_{x}\left(
g^{2}\partial_{x}%
{\displaystyle\sum\limits_{\text{even }n=0}^{\infty}}
\frac{\omega^{n}}{n!}\widetilde{X}^{(n)}\right)  \\
&  =g_{0}^{-1}%
{\displaystyle\sum\limits_{\text{even }n=2}^{\infty}}
\frac{\omega^{n}}{\left(  n-1\right)  !}\partial_{x}\widetilde{X}^{(n-1)}\\
&  =\omega^{2}g_{0}%
{\displaystyle\sum\limits_{\text{even }n=2}^{\infty}}
\frac{\omega^{n-2}}{\left(  n-2\right)  !}\widetilde{X}^{(n-2)}=\omega
^{2}u_{1}.
\end{align*}
In a similar way the solution $u_{2}$ can be verified as well.

\section{A remark on spectral problems}

First, let us notice that at least in the case of real-valued coefficients $p$
and $q$ under the conditions that $p$, $p^{\prime}$ and $q$ are continuous an
appropriate nonvanishing particular solution $g_{0}$ of (\ref{eqg0}) always
exists. In this case the equation $(pg^{\prime})^{\prime}+qg=0$ possesses two
linearly independent solutions $g_{1}$ and $g_{2}$ whose zeros do not
coincide. Then $g_{0}$ can be chosen as follows $g_{0}=g_{1}+ig_{2}$. Thus,
the proposed solution (\ref{genmain})-(\ref{genmain2}) can be obtained in a
quite general situation and as we show in this section the corresponding
spectral problems reduce to the problem of finding zeros of related analytic
functions. 

Sometimes it is slightly more convenient to consider the function
(\ref{genmain2}) divided by $\omega$, that is
\begin{equation}
u_{2}=g_{0}%
{\displaystyle\sum\limits_{\text{odd }n=1}^{\infty}}
\frac{\omega^{n-1}}{n!}X^{(n)}.\label{genmain3}%
\end{equation}
Then for the solutions of (\ref{Schrgen}) $u_{1}$ and $u_{2}$ defined by
(\ref{genmain1}) and (\ref{genmain3}) we obtain the following equalities%
\begin{equation}
u_{1}(0)=g_{0}(0),\qquad u_{1}^{\prime}(0)=g_{0}^{\prime}(0),\label{at01}%
\end{equation}%
\begin{equation}
u_{2}(0)=0,\qquad u_{2}^{\prime}(0)=\frac{1}{g_{0}(0)p(0)}.\label{at02}%
\end{equation}
Now consider a spectral problem for (\ref{Schrgen}) on the interval
$I_{x}=(0,1).$ For example,%
\begin{equation}
u(0)=0\quad\text{and\quad}u(1)=0.\label{bvpexample}%
\end{equation}
Due to the first boundary condition the constant $c_{1}$ in (\ref{genmain})
should be chosen as zero. Then the spectral problem reduces to finding such
values of $\omega$ that $u_{2}(1)=g_{0}(1)%
{\displaystyle\sum\limits_{\text{odd }n=1}^{\infty}}
\frac{\omega^{n-1}}{n!}X^{(n)}(1)$ vanish. In other words, this spectral
problem reduces to the calculation of zeros of the complex analytic function%
\[
\kappa(\omega)=%
{\displaystyle\sum\limits_{m=0}^{\infty}}
a_{m}\omega^{m}%
\]
where $a_{m}$ are defined as follows%
\[
a_{m}=\left\{
\begin{tabular}
[c]{ll}%
$0$ & $\text{for an odd }m$\\
$\frac{g_{0}(1)X^{(m+1)}(1)}{(m+1)!}$ & $\text{for an even }m.$%
\end{tabular}
\right.
\]
Note that in some cases the function $\kappa$ will be even entire. For
example, when the considered Sturm-Liouville problem is regular it is well
known (see, e.g., \cite{Levitan}) that there exists an unboundedly increasing
sequence of its eigenvalues which means that there exists an unboundedly
increasing sequence of zeros of the function $\kappa$.

Let $\alpha$ and $\beta$ be arbitrary real numbers. Consider the following
more general boundary conditions%
\begin{equation}
u(0)\cos\alpha+u^{\prime}(0)\sin\alpha=0\label{at0}%
\end{equation}%
\begin{equation}
u(a)\cos\beta+u^{\prime}(a)\sin\beta=0\label{atbeta}%
\end{equation}
together with equation (\ref{Schrgen}). Taking the solutions $u_{1}$ and
$u_{2}$ defined by (\ref{genmain1}) and (\ref{genmain3}) and using
(\ref{at01}), (\ref{at02}) we obtain from (\ref{at0}) the following equation
for $c_{1}$ and $c_{2}$
\[
c_{1}(g_{0}(0)\cos\alpha+g_{0}^{\prime}(0)\sin\alpha)+c_{2}\frac{\sin\alpha
}{g_{0}(0)p(0)}=0
\]
which gives the relation
\[
c_{2}=\gamma c_{1}\quad\text{for }\alpha\neq\pi n
\]
where $\gamma=-g_{0}(0)p(0)(g_{0}(0)\cot\alpha+g_{0}^{\prime}(0))$ and
\[
c_{1}=0\quad\text{for }\alpha=\pi n.
\]
In this last case we arrive at a similar result as in the example considered
above, thus let us consider the case $\alpha\neq\pi n$. From the definition of
$u_{1}$ and $u_{2}$ we have%
\[
u_{1}^{\prime}=\frac{g_{0}^{\prime}}{g_{0}}u_{1}+\frac{1}{g_{0}p}%
{\displaystyle\sum\limits_{\text{even }n=2}^{\infty}}
\frac{\omega^{n}}{\left(  n-1\right)  !}\widetilde{X}^{(n-1)}%
\]
and
\[
u_{2}^{\prime}=\frac{g_{0}^{\prime}}{g_{0}}u_{2}+\frac{1}{g_{0}p}%
{\displaystyle\sum\limits_{\text{even }n=0}^{\infty}}
\frac{\omega^{n}}{n!}X^{(n)}.
\]
Then the boundary condition (\ref{atbeta}) implies the following equation%
\[
\left(  g_{0}(a)\cos\beta+g_{0}^{\prime}(a)\sin\beta\right)  \left(
{\displaystyle\sum\limits_{\text{even }n=0}^{\infty}}
\frac{\omega^{n}}{n!}\widetilde{X}^{(n)}(a)+\gamma%
{\displaystyle\sum\limits_{\text{odd }n=1}^{\infty}}
\frac{\omega^{n-1}}{n!}X^{(n)}(a)\right)
\]%
\[
+\frac{\sin\beta}{g_{0}(a)p(a)}\left(
{\displaystyle\sum\limits_{\text{even }n=2}^{\infty}}
\frac{\omega^{n}}{\left(  n-1\right)  !}\widetilde{X}^{(n-1)}+\gamma%
{\displaystyle\sum\limits_{\text{even }n=0}^{\infty}}
\frac{\omega^{n}}{n!}X^{(n)}\right)  =0.
\]
Thus the spectral problem (\ref{Schrgen}), (\ref{at0}), (\ref{atbeta}) reduces
to the problem of finding zeros of the analytic function%
\[
\kappa(\omega)=%
{\displaystyle\sum\limits_{m=0}^{\infty}}
a_{m}\omega^{m}%
\]
where $a_{m}$ are defined as follows
\[
a_{0}=\left(  g_{0}(a)\cos\beta+g_{0}^{\prime}(a)\sin\beta\right)  (1+\gamma
X^{(1)}(a))+\frac{\gamma\sin\beta}{g_{0}(a)p(a)}%
\]
and%
\[
a_{m}=\left\{
\begin{tabular}
[c]{ll}%
$0$ & $\text{for an odd }m$\\
& \\
$\left(  g_{0}(a)\cos\beta+g_{0}^{\prime}(a)\sin\beta\right)  \left(
\frac{\widetilde{X}^{(m)}(a)}{m!}+\gamma\frac{X^{(m+1)}(a)}{\left(
m+1\right)  !}\right)  $ & \\
$+\frac{\sin\beta}{g_{0}(a)p(a)}\left(  \frac{\widetilde{X}^{(m-1)}%
(a)}{\left(  m-1\right)  !}+\gamma\frac{X^{(m)}(a)}{m!}\right)  $ & $\text{for
an even }m>0.$%
\end{tabular}
\right.
\]

\section{A remark on the Darboux transformation}

The Darboux transformation is a very useful and important tool studied in
dozens of works (see, e.g., \cite{MS}). It is closely related to the
factorization of the Schr\"{o}dinger operator (\ref{fact}). Consider the
equation
\[
\left(  \partial_{x}+\frac{g_{0}^{\prime}}{g_{0}}\right)  \left(  \partial
_{x}-\frac{g_{0}^{\prime}}{g_{0}}\right)  u=\omega^{2}u.
\]
Applying the operator $\left(  \partial_{x}-\frac{g_{0}^{\prime}}{g_{0}%
}\right)  $ to both sides and denoting $v=\left(  \partial_{x}-\frac
{g_{0}^{\prime}}{g_{0}}\right)  u$ one obtains that solutions of equation
(\ref{Schrmain}) are transformed into solutions of another Schr\"{o}dinger
equation%
\[
\left(  \partial_{x}-\frac{g_{0}^{\prime}}{g_{0}}\right)  \left(  \partial
_{x}+\frac{g_{0}^{\prime}}{g_{0}}\right)  v=\omega^{2}v
\]
which can be written also as follows%
\begin{equation}
\left(  -\partial_{x}^{2}+r(x)+\omega^{2}\right)  v(x)=0, \label{SchrDarboux}%
\end{equation}
where $r=2\left(  \frac{g_{0}^{\prime}}{g_{0}}\right)  ^{2}-q$. Now, as we are
able to construct the general solution of (\ref{Schrmain}) by a known solution
of (\ref{Schrwithout}) we can also obtain an explicit form of the result of
the Darboux transformation. First, let us apply the operator $\left(
\partial_{x}-\frac{g_{0}^{\prime}}{g_{0}}\right)  =g_{0}\partial_{x}g_{0}%
^{-1}$ to $u_{1}$ defined by (\ref{gensol1}). We have
\begin{align*}
v_{1}  &  =\left(  \partial_{x}-\frac{g_{0}^{\prime}}{g_{0}}\right)
u_{1}=g_{0}%
{\displaystyle\sum\limits_{\text{even }n=0}^{\infty}}
\frac{\omega^{n}}{n!}\partial_{x}\widetilde{X}^{(n)}\\
&  =g_{0}^{-1}%
{\displaystyle\sum\limits_{\text{even }n=2}^{\infty}}
\frac{\omega^{n}}{\left(  n-1\right)  !}\widetilde{X}^{(n-1)}=\frac{\omega
}{g_{0}}%
{\displaystyle\sum\limits_{\text{odd }n=1}^{\infty}}
\frac{\omega^{n}}{n!}\widetilde{X}^{(n)}%
\end{align*}
and in a similar way we obtain
\[
v_{2}=\left(  \partial_{x}-\frac{g_{0}^{\prime}}{g_{0}}\right)  u_{2}%
=\frac{\omega}{g_{0}}%
{\displaystyle\sum\limits_{\text{even }n=0}^{\infty}}
\frac{\omega^{n}}{n!}X^{(n)}.
\]
Thus, the general solution of the Schr\"{o}dinger equation (\ref{SchrDarboux})
obtained from (\ref{Schrmain}) by the Darboux transformation has the form%
\[
v=\frac{c_{1}}{g_{0}}%
{\displaystyle\sum\limits_{\text{even }n=0}^{\infty}}
\frac{\omega^{n}}{n!}X^{(n)}+\frac{c_{2}}{g_{0}}%
{\displaystyle\sum\limits_{\text{odd }n=1}^{\infty}}
\frac{\omega^{n}}{n!}\widetilde{X}^{(n)}%
\]
where $X^{(n)}$ and $\widetilde{X}^{(n)}$ are defined by (\ref{X1})-(\ref{X3}).

\section{Conclusions}

The main result of this work allows us to find the general solution of
(\ref{main}) by a known solution of (\ref{main0}) as follows. By formulas
(\ref{Xgen1})-(\ref{Xgen3}) the functions $X^{(n)}$ and $\widetilde{X}^{(n)}$
should be constructed and then the general solution of (\ref{main}) has the
form (\ref{genmain})-(\ref{genmain2}). An important feature of these formulas
consists in the fact that the form of $X^{(n)}$ and $\widetilde{X}^{(n)}$ does
not depend on $\omega$. For given coefficients $p$ and $q$ they should be
calculated only once and then the general solution for any $\omega$ is
obtained by multiplying them by corresponding powers of $\omega$. This is very
convenient for numerical calculations. Moreover, our numerical experiments
show that $X^{(n)}$ and $\widetilde{X}^{(n)}$ can be calculated up to high
indices with a remarkable accuracy, thus we expect that among other possible
applications the results of this work will be useful in numerical solving of a
wide class of boundary value and spectral problems of mathematical physics.

\textbf{Acknowledgement}

The author wishes to express his gratitude to CONACYT for supporting this work
via the research project 50424.

\end{document}